\documentclass[showpacs,epsf,onecolumn]{revtex4-1}
\usepackage{float}
\usepackage{amssymb}
\usepackage{color,graphicx}
\usepackage{hyperref}
\begin{document}
\title{Atoms to topological electronic materials: A bedtime story for beginners}
\author{Arnab Kumar Pariari}
\email{Correspondence to ``arnab.pariari@weizmann.ac.il"}
\begin{abstract}
In this review, We discussed the theoretical foundation and experimental discovery of different topological electronic states of material in condensed matter. At first, we briefly reviewed the conventional electronic states, which have been realized in band theory of solid. Next, the simplest non-trivial insulating phase (Integer Quantum Hall State) and the concept of topological order in condensed matter electronic system have been introduced. In the following sections, we discussed Quantum Spin Hall (QSH) State in two dimensions (2D), and reviewed the theoretical and experimental developments from 2D QSH state to 3D topological insulators (TI). Subsequently, we gave a brief overview on theoretical and experimental understanding on recently discovered topological Dirac semimetals, Weyl semimetals, three-, six- , and eight-fold degenerate semimetals, and Nodal line semimetals.
Then, topological crystalline insulator, which can not be considered as a descendent of Quantum Spin Hall or Integer Quantum Hall insulator, has been introduced. Finally, we discussed the presence of magnetism in some topological materials and its consequence on electronic band structure.
\begin{figure}[h]
\includegraphics[width=0.8\textwidth]{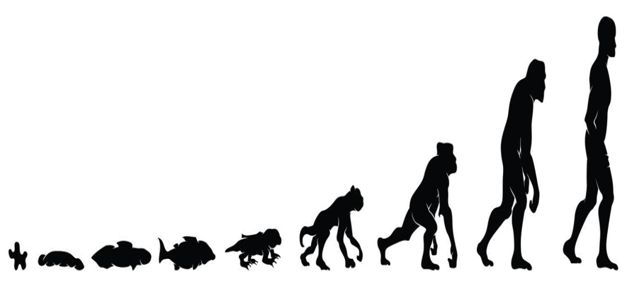}
\end{figure}
\end{abstract}
\maketitle

\section{Band theory and conventional electronic phases}\label{PO}
\begin{figure}[h]
\includegraphics[width=0.8\textwidth]{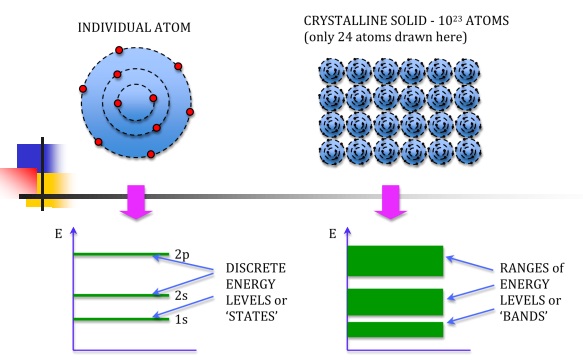}
\caption{A schematic diagram to show the discrete energy levels of an isolated atom and energy band of crystalline solid. Reproduced from Ref. \cite{1}.}\label{rh}
\end{figure}

\begin{figure}[h]
\includegraphics[width=0.8\textwidth]{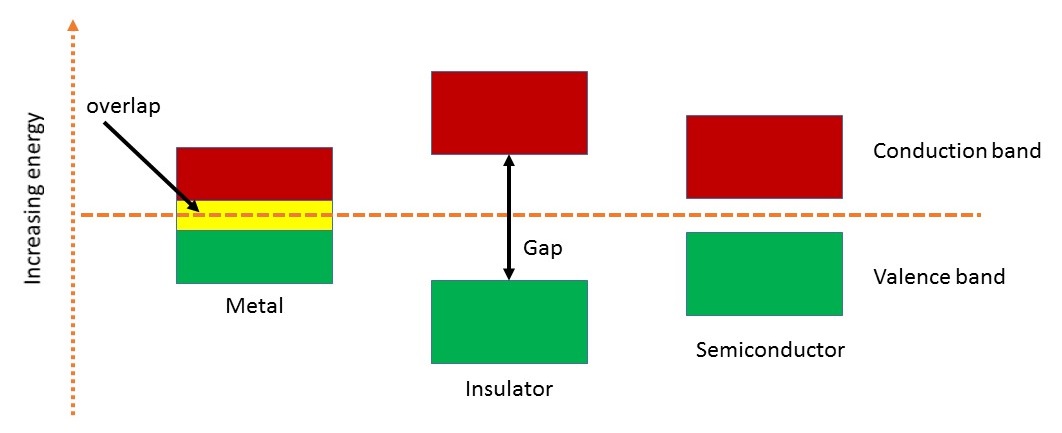}
\caption{Electronic phases of matter classified by band theory.}\label{rh}
\end{figure}
In the case of a single isolated atom, there are various discrete energy levels, known as atomic orbitals. When two atoms join together to form a molecule, their atomic orbitals overlap, and each atomic orbital splits into two molecular orbitals of different energy. In a solid, a large number of atoms are arranged systematically in space lattice and each atom is influenced by neighbouring atoms. As a consequence, each atomic orbital splits into large number of discrete molecular orbitals, each with a different energy. The energy of adjacent levels is so close that they can be considered as a continuum, forming an energy band. Figure 1 is a schematic diagram, representing the above discussion. The highest completely occupied band is called the valence band and the partially filled or completely empty band is known as conduction band. For the conduction of electrical energy in a material, there must be partially filled band. In case of a metal, as shown in Figure 2, the valence and conduction band overlap with each other in such a way that the conduction band is partially filled and participates in charge conduction. A semimetal, where the valence and conduction band just touch at a point without introducing any well-defined Fermi surface, is also a conductor of charge. For an insulator, due to gap between valence and conduction band, the conduction band is completely empty and there is no charge conduction under external electric field. In the materials where the gap is small ($\lesssim$ 1 eV), electrons thermally excited from valence to conduction band near room temperature, and participate in charge conduction. These materials are known as semiconductors. It can be shown that via smooth deformation of the Hamiltonian, an insulating gap can be tuned to an arbitrarily small value or to an exceptionally large value, without closing the gap. In mathematical language, all the conventional insulating states are related via an equivalence relation. In that sense, vacuum, which, according to the Dirac equation, has a band gap that corresponds to the pair production ($\sim$ 10$^{6}$ eV), is also a trivial insulator. Thus, the band theory of solid is extremely successful in grouping a wide variety of materials into just two categories: metals and insulators. It has been thought to be most powerful quantum mechanical tool available to understand the electronic properties of crystalline solids, until the discovery of Integer Quantum Hall Effect (IQHE).
\section{The Integer Quantum Hall State and introduction of topology in electronic systems}\label{PO}
\subsection{Experimental discovery and the strange observation}\label{PO}
\begin{figure}[h]
\includegraphics[width=1.0\textwidth]{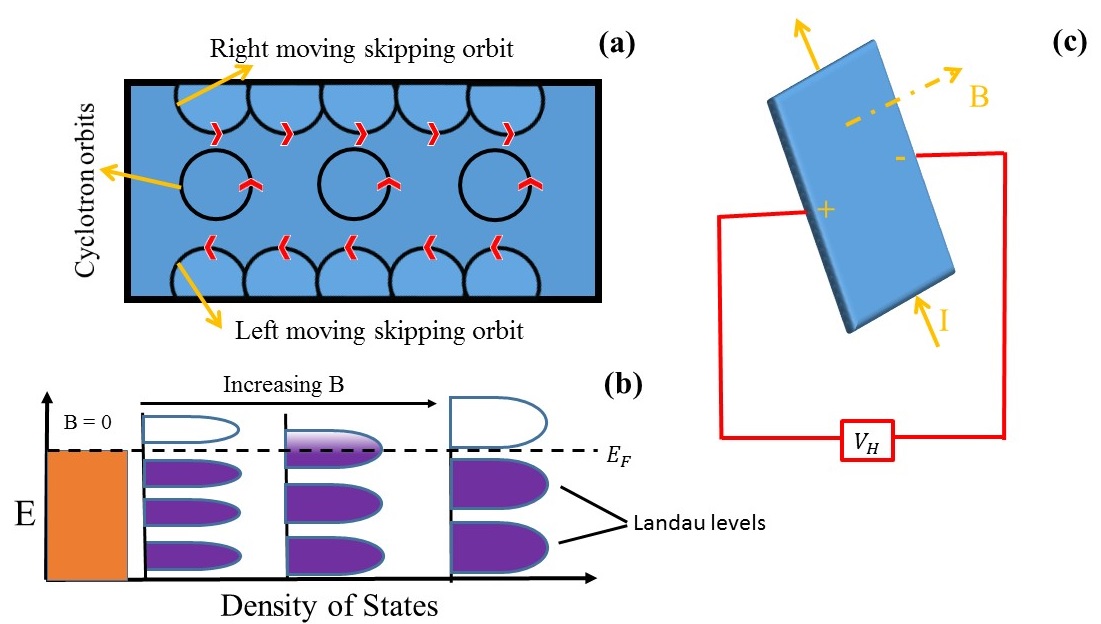}
\caption{(a) Schematic of two-dimensional electron gas under out-of-plane external magnetic field. (b) Formation of Landau levels under application of magnetic field, and the variation in the electronic density of states across the Fermi level with the increasing field. (c) Hall measurement configuration. V$_{H}$ is the Hall voltage.}\label{rh}
\end{figure}
The experimental discovery of Integer Quantum Hall Effect in 1980 by von Klitzing, led to think a different classification paradigm, beyond conventional band theory of solid \cite{2,3}. IQHE is the simplest example of insulator which is fundamentally not equivalent to vacuum. Two-dimensional electron gas under application of out-of-plane external magnetic field forms cyclotron orbits well-inside the boundary [Figure 3(a)]. The single particle Hamiltonian ($H$) describing the motion of electron is given by the expression,
\begin{equation}
H = \frac {(\textbf{p} - e\textbf{A})^{2}}{2m},
\end{equation}
where $p$, $m$, and $\textbf{A}$ are momentum of electron, effective mass, and magnetic vector potential, respectively. In this situation, each electronic energy band of parent state splits into several sub-bands, known as Landau levels. The energy of n$^{th}$ Landau level is $E_n = \hbar \omega_c(n + \frac {1}{2})$, where $\omega_{c} = eB/m^{\star}_{c}$ and $m^{\star}_{c}$ is the effective cyclotron mass of the charge carrier. When N Landau levels are filled, there is an energy gap between N$^{th}$ filled band and (N+1)$^{th}$ empty band, which causes the bulk to behave as an insulator. Being a function of external magnetic field, the degeneracy of Landau levels increases with increasing field strength. As a consequence, the Landau levels pass through the Fermi level of the system, which results in oscillations of the electronic density of states at the Fermi level. This phenomenon produces oscillations in several electronic properties of a material including electrical resistance (Shubnikov-de Haas effect) and magnetization (de Haas-van Alphen effect), which is familiar as quantum oscillations. The frequency of this oscillation in a material is proportional to the cross-sectional area of the Fermi surface, perpendicular to the direction of magnetic field. By applying magnetic field along different directions of a crystals, one can measure the cross sections of the Fermi surface. This technique has been established as a powerful tool to probe the Fermi surface of a material.\\

However, the electrons at the edge of the two-dimensional electron gas will behave differently from that of the bulk, as shown in Figure 3(a). Due to the bending of the path by the Lorentz force, electrons form skipping orbits. Hall conductivity ($\sigma_{xy}$), which has been obtained by measuring the Hall resistivity as shown in Figure 3(c), is found to be finite, unlike trivial insulators, and $\sigma_{xy}$ is quantized depending on the number ($N$) of filled Landau levels. The quantized value of Hall conductivity is given by the expression, $\sigma_{xy} = \frac {N e^{2}}{h}$ \cite{2,4}. The mysterious thing about the value of $\sigma_{xy}$ is that the quantization can be measured to an accuracy 1 part in a billion. Irrespective of material forming the two-dimensional electron system and presence of disorder, which modify the Hamiltonian of the system, the value of $\sigma_{xy}$ has been found to have such precise quantization. To explain the robust and quantized value of $\sigma_{xy}$, concept of topological order has been introduced in solid state electronic systems.\\
\subsection{Topology in Geometry}\label{PO}
\begin{figure}[h]
\includegraphics[width=0.8\textwidth]{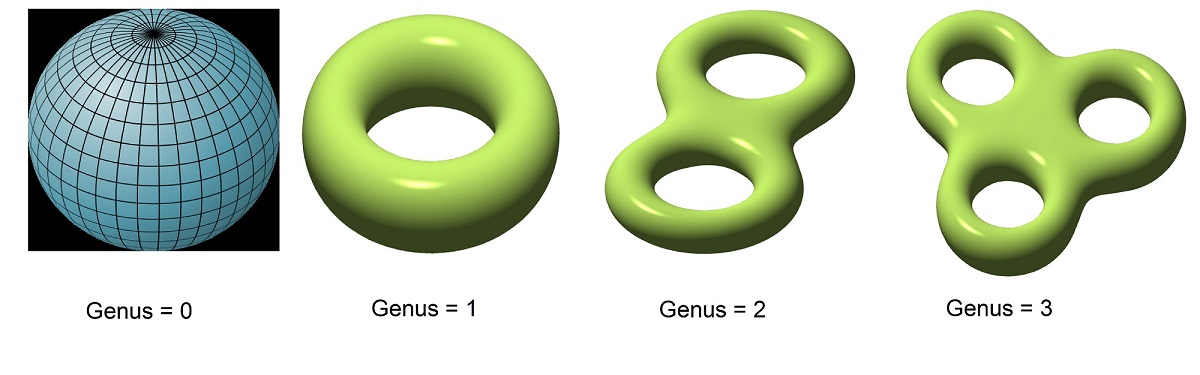}
\caption{Geometrical objects with different topology. The objects are classified by the value of Genus, which is basically the number of holes in the object.}\label{rh}
\end{figure}
Topology is a mathematical structure in Geometry, and this allows us to study the properties of an object, which remain unaffected by the smooth deformation of shape or size. In Figure 4, four three-dimensional objects have been shown, which belong to different topological class. The topological quantity which distinguishes a sphere from a torus, is called genus ($g$). The values of genus for sphere, torus, double torus, and triple torus are zero, one, two, and three, respectively. So, the value of $g$ is basically the number of holes in an object. Since an integer can not change smoothly, objects with different genus can not be deformed into one another, and are said to be topologically distinct. In that sense, a clay ball and a plate, both of which have $g$ = 0, can be deformed smoothly into one another. In other words, any two objects with the same value of $g$ can be connected by a smooth deformation in size or shape. The genus of an object with arbitrary shape is calculated by the Gauss-Bonnet theorem,
\begin{equation}
\int_{Surface} K dS = 2\pi(2 - 2g),
\end{equation}
where $K = \frac {1}{r_{1}r_{2}}$ is the Gaussian curvature, and r$_{1}$ and r$_{2}$ are the radius along two perpendicular directions from a point on the surface of an object \cite{2,5}. Considering, $K = \frac {1}{r^{2}}$ on the surface of a sphere, $\int_{Surface} K dS$ has been calculated to be 4$\pi$. This implies that the value of $g$ for a sphere is zero. If we perform similar calculation for a plate like object, it will give us the same value for $g$.\\

\subsection{Topology in Quantum Hall Physics}\label{PO}
How the concept of topology can be used to characterize Integer Quantum Hall states? We will explain this in this paragraph. From the mathematical point of the view, the Gaussian curvature of geometry, the Berry curvature of electronic band theory and magnetic field are same. All of them are described by the same mathematical structure: fiber bundles \cite{5}. Now the question is: How the Berry's phase arises in solid state electronic systems? The band theory of solid classifies electronic states in terms of their crystal momentum \textbf{k}, defined in a periodic Brillouin zone. The Bloch states $|u_{m}(\textbf{k})>$, defined in a single unit cell of the crystal, are eigenstates of the Bloch Hamiltonian, $H(\textbf{k})$. The eigenvalues $E_{m}(\textbf{k})$ for all $m$, collectively form the band structure. However, the Bloch wave function, $|u_{m}(\textbf{k})>$, has an intrinsic phase ambiguity, $e^{i\phi(\textbf{k})}$. The band structure remains unaffected under the transformation, $|u_{m}(\textbf{k})> \longrightarrow e^{i\phi(\textbf{k})}|u_{m}(\textbf{k})>$, which is similar to gauge transformation in electromagnetic theory. This leads to introduce a quantity similar to electromagnetic vector potential, which transforms $\textbf{A}_{m} \longrightarrow \textbf{A}_{m} + \nabla_{\textbf{k}}\phi(\textbf{k})$ under gauge transformation. So, there must be an analog of magnetic flux, $F_{m} = \nabla_{\textbf{k}}\times\textbf{A}_{m}$, which is invariant under the transformation \cite{2}. This quantity is known as Berry curvature, and $\textbf{A}_{m}$ is defined as $\textbf{A}_{m} = i<u_{m}|\nabla_{\textbf{k}}|u_{m}>$. Thouless, Kohmoto, Nightingale, and den Nijs have found that the surface integral of Berry curvature over the Brillouin zone is an integer, $\int_{B.Z.} F_{m} d^{2}k = n_{m}$, similar to genus in geometry \cite{2,6}. The topological invariant $n_{m}$ is called Chern invariant, and the total Chern number, summed over all occupied bands, $n = \sum_{m=1}^{N} \in \mathbb{Z}$ ($\mathbb{Z}$ denotes the integer, i.e., 1, 2, ....., $\infty$) is invariant, provided the gap separating occupied and empty bands remains finite. $n$ is also known as TKNN invariant. It has been identified that this $n$ is nothing but the integer number in the expression, $\sigma_{xy} = \frac {N e^{2}}{h}$ \cite{6}. Being a topological invariant, $n$ in a system can not be changed under smooth deformation of Hamiltonian, i.e., without closing the gap between the occupied and empty bands. This helps us to explain the robust quantization of $\sigma_{xy}$ in quantum Hall state.\\

\begin{figure}[h]
\includegraphics[width=0.8\textwidth]{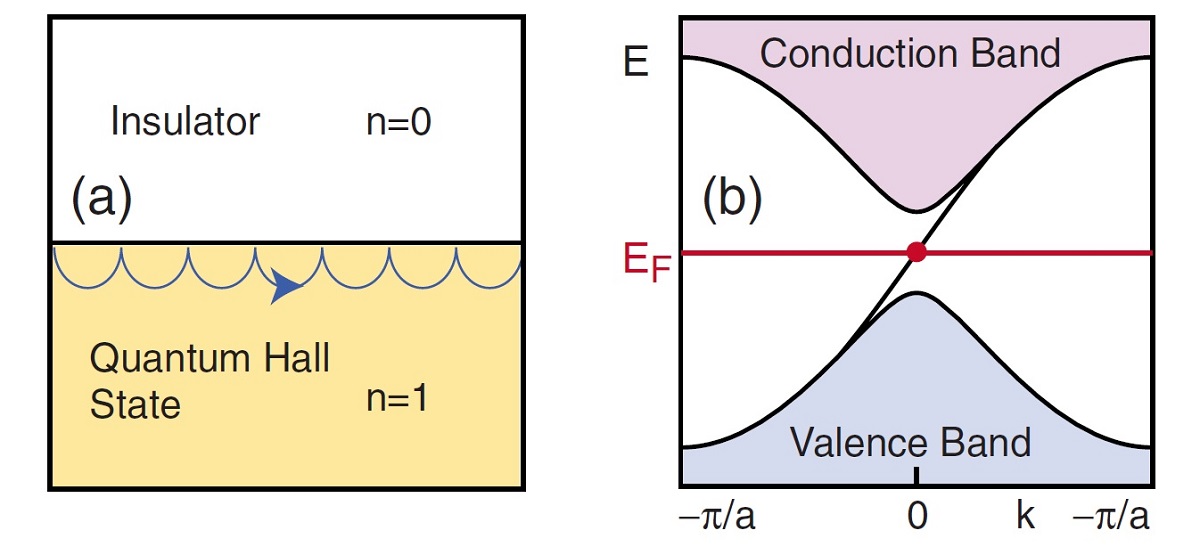}
\caption{(a) The interface between a quantum Hall ground state and an trivial insulator/vacuum. (b) The electronic band structure, where a single edge state connects the bulk valence band to the bulk conduction band. Reproduced from Ref. \cite{2}.}\label{rh}
\end{figure}
The existence of skipping electron orbits or metallic edge state at the interface of Quantum Hall state and vacuum is the fundamental consequence of the topological classification in gapped states. Topological protection prevents states to deform smoothly from one value of $n$ to another, across the interface of two topologically different insulators. As shown in Figure 5(a), the Quantum Hall ground state has the value of $n$ equals to one \cite{2}. Whereas, a trivial insulator/vacuum has $n$ equals to zero. As a consequence, the Hamiltonian can not be smoothly deformed from Integer Quantum Hall state to trivial insulating state. The gap between the valence and conduction band must be close to change the value of $n$ at the boundary. This provides an one-dimensional band dispersion for the edge state, residing in the bulk band gap [Figure 5(b)] \cite{2}. The number of edge channels at the interface of two topologically different systems is determined by the `bulk-boundary correspondence' \cite{2}. This relates the number of edge modes ($N$) intersecting the Fermi energy to the change in the bulk topological invariant ($n$) across the interface by the expression, $N = \triangle n$. So, a Quantum Hall state with $N$ number of filled Landau levels will have $N$ number of edge channels at the interface with vacuum.\\
\section{Quantum Spin Hall (QSH) State}\label{PO}
\subsection{Discovery of QSH effect and failure of TKNN characterization}\label{PO}
\begin{figure}[h]
\includegraphics[width=0.8\textwidth]{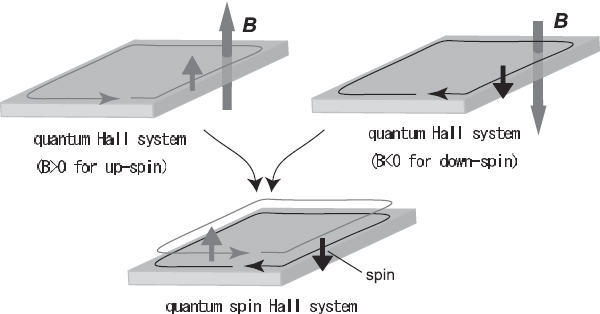}
\caption{Schematic picture of the QSH system as a superposition of two QH systems. Reproduced from Ref. \cite{8}.}\label{rh}
\end{figure}
Discovery of Quantum spin Hall insulator state in HgTe quantum-well by Molenkamp and his collaborators, in the year 2007, is the next milestone in classifying electronic states of matter in terms of their underlying topology \cite{7}. The two-dimensional quantum-well structure was made by sandwiching a thin layer of mercury telluride (HgTe) between layers of cadmium telluride (CdTe). In QSH state, time reversal symmetry is preserved due to absence of external magnetic field and spin-orbit coupling plays an important role in generating intrinsic magnetic field, unlike IQH state. The single particle effective Hamiltonian, governing the Quantum Spin Hall physics is,
\begin{equation}
H = \frac {(\textbf{p} - e\textbf{A}\sigma_{z})^{2}}{2m},
\end{equation}
where $\sigma_{z}$ is the z-component of Pauli matrices. It is evident from the second term (within the bracket) in the numerator that an effective magnetic field acts in the upward direction on up-spin and in the downward direction on down-spin. As a result, electrons with upward spin move in a separate conducting channel, opposite to the spin-down electrons, at the edge of the sample. So, a QSH phase can be realized by a superposition of two quantum Hall systems for the up and down spins, as shown in Figure 6 \cite{8}. However, there is no net flow of charge, but net spin current in QSH state. A QSH insulator can not be characterized by the TKNN invariant ($n\in Z$). This is because the integer topological invariant for up-spin electrons ($n\uparrow$) is equal and opposite to the down-spin electrons ($n\downarrow$) in presence of time reversal symmetry, and as a consequence, $n$ ( =$n\uparrow$ + $n\downarrow$) is zero. Considering the role of spin-orbit interaction and time-reversal ($\mathcal{T}$ ) symmetry, Kane, Mele, and others have introduced a new topological invariant, $\nu$ \cite{9,10}.\\

\subsection{Role of time-reversal symmetry}\label{PO}
To understand this new topological class, we have to examine the role of $\mathcal{T}$  symmetry for spin-$\frac{1}{2}$ particles. The $\mathcal{T}$ symmetry in an arbitrary spin system is represented by an anti-unitary operator, $\Theta = exp(\frac{i}{\hbar} \pi S_y) K$, where $S_y$ is the spin operator and $K$ is the complex conjugate. Existence of time-reversal symmetry implies that $\Theta$ commutes with the Hamiltonian of the system (where $\Psi$ represents the wave function of the system), i.e.,\\
$[\Theta, H]\Psi = 0$\\
$\Rightarrow \Theta H \Psi - H \Theta \Psi = 0$\\
$\Rightarrow \Theta H \Psi = H \Theta \Psi$\\
Let, $\Psi$ is the r$^{th}$ eigen state of $H$, i.e., $H \Psi = \varepsilon_{r}\Psi$. This implies, $H \Theta \Psi = \varepsilon_{r} \Theta \Psi$. So, $\Theta \Psi$ is also the r$^{th}$ eigen state of $H$. Now, there are two possibilities: (a) $\Psi$ and $\Theta \Psi$ are same, and (b) $\Psi$ and $\Theta \Psi$ are different wave functions, i.e., $\varepsilon_{r}$ is doubly degenerate. To identify the right one, we have to consider the following effect of $\mathcal{T}$  symmetry operation on spin-half system. In a spin-half system, $\Theta$ flips the direction of the spin by 180$^{0}$ and wave function gains a minus sign by the two times operation of $\Theta$, i.e., $\Theta^{2} \Psi = -\Psi$.\\
Suppose, $\Psi = \Theta \Psi$\\
$\Rightarrow \Theta^{2} \Psi = \Theta \Psi = \Psi$\\
$\Rightarrow \Psi \neq \Theta \Psi$\\
So, condition (b) is right, which states $\Psi$ and $\Theta \Psi$ are independent wave functions, i.e., $\varepsilon_{r}$ is doubly degenerate. This is the famous Kramer's theorem, which states that ``all eigenstates of a $\mathcal{T}$-invariant Hamiltonian of spin-half system are twofold degenerate".\\

\begin{figure}[h]
\includegraphics[width=1.0\textwidth]{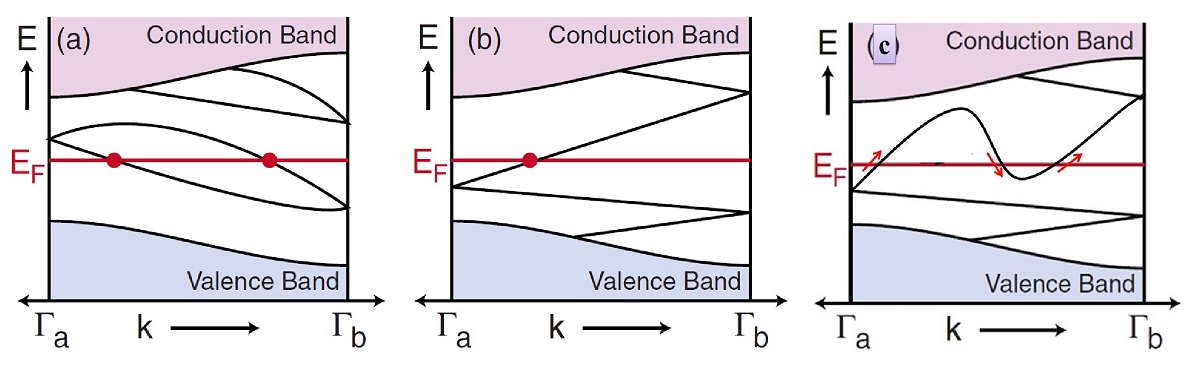}
\caption{(a) The edge states cross the Fermi level an even (zero) number of times. (b) The edge states cross the Fermi level once. (c) The edge states cross the Fermi level an odd number of times.  Reproduced from Ref. \cite{2}.}\label{rh}
\end{figure}
If the Kramer's theorem is applied for the Bloch wave state of solid, it will be found that for any Bloch wave state $\Psi_{k}$, there is another state $\Theta \Psi_{\textbf{k}}$ = $\Psi_{-\textbf{k}}$ of same energy. So, Kramer's doublets are located at different momentum point \textbf{k} and -\textbf{k}. Only at \textbf{k} = 0 and \textbf{k} = $\pi$ (considering lattice parameter is a unit quantity), both the points are the same. This implies that at \textbf{k} = 0 and \textbf{k} = $\pi$, each Bloch state comes in pair. On the other hand, single particle Hamiltonian of an electronic system smoothly deforms from the bulk to edges. If any edge state is induced inside the bulk band gap, at \textbf{k} = 0 and \textbf{k} = $\pi$, it will be doubly degenerate. Away from these special points, the spin-orbit interaction will split the degeneracy. As electronic band dispersion is continuous, the states at \textbf{k} = 0 and \textbf{k} = $\pi$ have to be connected. But there is only two possible ways [Figure 7], through which they can connect. For the first case [Figure 7 (a)], edge state crosses the Fermi level at an even (zero) number of points. So, there will be even numbers of conducting channels or no channel at the edge. In this case, the edge states can be eliminated by tuning the Fermi level, or by smooth deformation of Hamiltonian in such a way that all the Kramer's doublets appear outside the gap. In conclusion, pairwise interconnection of states at \textbf{k} = 0 and \textbf{k} = $\pi$ gives rise to trivial insulating phase. For the second case [Figure 7(b)], when the edge band crosses the Fermi level once, there is single conducting edge channel. This type of edge state is unavoidable under any smooth deformation of Hamiltonian or shifting of Fermi level. In this context, one can suggest the third possibility [Figure 7(c)], where the edge band crosses the Fermi levels three times. However, this type of connection will generate two right-moving and one left-moving channel, and as a consequence, there will be an effective single conducting edge state. The one-to-one connection of states at \textbf{k} = 0 and \textbf{k} = $\pi$ , as shown in Figure 7(b) and Figure 7(c), leads to  topologically protected metallic boundary states. Which of the above-mentioned scenarios will occur at the edge, will be determined by the topological class of the bulk band structure?\\

\subsection{$\mathbb{Z}_{2}$ topological classification}\label{PO}
\begin{figure}[h]
\includegraphics[width=1.0\textwidth]{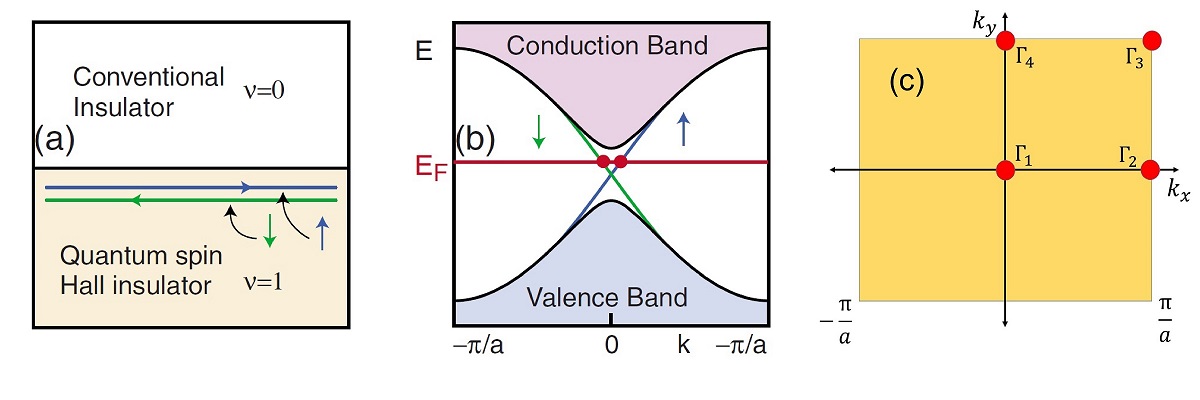}
\caption{(a) The interface between a QSH insulator and an ordinary insulator.(b) The edge state dispersion. (c) High-symmetry points in 2D bulk Brillouin zone. Figure (a) and (b) are reproduced from Ref. \cite{2}.}\label{rh}
\end{figure}
According to the `bulk-boundary correspondence' principle, the number of edge modes has to be equal to the change in the bulk topological invariant ($\triangle\nu$) across the interface. This implies that the bulk topological invariant ($\nu$) for two-dimensional insulating states in presence of time-reversal symmetry, has to be either zero or one. $\nu$ obeys all the group operations of two-dimensional cyclic group ($\mathbb{Z}_{2}$) such as, \textbf{a} $\oplus$ \textbf{a} = \textbf{a}, \textbf{a}$\oplus$ \textbf{b}= \textbf{b} and \textbf{b}$\oplus$ \textbf{b}= \textbf{a}, where \textbf{a} and \textbf{b} are the two group elements, and \textbf{a} is the unity element of $\mathbb{Z}_{2}$. As a consequence, $\nu$ has been named as `$\mathbb{Z}_{2}$ topological invariant'. As shown in Figure 8(a), for Quantum Spin Hall insulator, $\nu$ encounters a unit change across the interface with trivial insulator/vacuum. This leads to single edge state for spin-up electron state and spin-down electron state, separately. Figure 8(b) shows the corresponding band structure in first Brillouin zone. Now the question is: ``how the value of $\nu$ is determined for a two-dimensional insulating state in presence of time-reversal symmetry?" There are several mathematical formalisms for determining the value of $\nu$, however, the method, which has been developed by Fu and Kane, will be mentioned here \cite{11}. In order to calculate $ \nu $, the authors have initially defined a unitary matrix $w_{mn}(\textbf{k}) = <u_m(\textbf{k})\mid {\Theta}\mid u_n (\textbf{k})>$, using the occupied Bloch functions. As $\Theta^2 = -1$, it can be shown that $w^T (\textbf{k}) = - w (-\textbf{k})$. For a two-dimensional electronic system, there are four inequivalent special points in the bulk Brillouin zone, which have been identified as $\Gamma_{a = 1, 2, 3, 4}$, in Figure 8(c). In these points, \textbf{k} and -\textbf{k} are equivalent, which makes $ w(\Gamma_a)$ antisymmetric matrix. The determinant of an antisymmetric matrix is the square of its pfaffian, which allows us to define a quantity, $\delta_a = \frac{\mathrm{Pf}[w(\Gamma_a)] }{\sqrt{\det[w(\Gamma_a)]}} = \pm 1$. The invariant $\nu $ is determined by the expression,
\begin{equation}
( -1) ^{\nu} = \prod_{a = 1}^{4} \delta_a.
\end{equation}

\section{Three-dimensional Topological Insulators}\label{PO}
Shortly after the discovery of QSH insulator in two-dimension, its three-dimensional (3D) counterpart has been realized theoretically, which has been named as ` 3D topological insulator (3D TI)' \cite{12,13}. Similar to the conducting edge state of 2D-QSHI, 3D TI has topology protected surface state which crosses the Fermi level, residing in the bulk band gap. Time-reversal invariant 3D bulk insulating state has also been characterized by $\mathbb{Z}_{2}$ topological invariant. However, 3D topological insulators are described by four $\mathbb{Z}_{2}$ topological invariants ($\nu_0$;$\nu_1$$\nu_2$$\nu_3$), instead of single invariant in two dimensions. $\nu_0$ is known as strong topological index, and the other three are known as weak topological indices. It is customary to write the combination of the four invariants in the form ($\nu_0$;$\nu_1$$\nu_2$$\nu_3$), because ($\nu_1$$\nu_2$$\nu_3$) can be interpreted as Miller indices to specify the direction of vector $\Gamma_a$ in the reciprocal space. In the following section, we will discuss two types of 3D TI state, depending on the value of $\nu_0$ \cite{2,3}.\\
\subsection{Weak topological insulators}\label{PO}
$\nu_0$ = 0 represents the simplest 3D TI, which can be understood by stacking the layers of QSHI, with weak interlayer coupling. The orientation of layers is described by ($\nu_1$$\nu_2$$\nu_3$) such as, (0 0 1) represents stacking along \textbf{z}-axis. The conducting edge-state of monolayers, as shown by the blue arrows in the Figure 9, forms a topological surface state in bulk sample. A simple cubic Brillouin zone for the three-dimensional bulk electronic system has eight time-reversal invariant points, which have been shown by red dots in Figure 9(b). Each of the planes in the Brillouin zone (e.g., $k_i = 0, \pi$ planes) containing four such points, is characterized by a 2D invariant, which is calculated using the equation,
\begin{equation}
( -1) ^{\nu_{i = 1, 2, 3}} = \prod_{a = 1}^{4} \delta_a.
\end{equation}
Earlier, it was believed that the surface state is present for a clean sample of a weak TI (WTI), but in presence of disorder, it can be localized. Later on, the surface states of WTI are found to be protected from random impurities and disorders, which do not break the time-reversal symmetry and close the bulk energy gap \cite{14}. As a consequence, the surface conductance of a WTI remains finite even in presence of strong disorder. $Bi_{14}Rh_{3}I_{9}$ is one of the compounds, which has been experimentally addressed to be a weak TI \cite{15}.\\
\begin{figure}[h]
\includegraphics[width=1.0\textwidth]{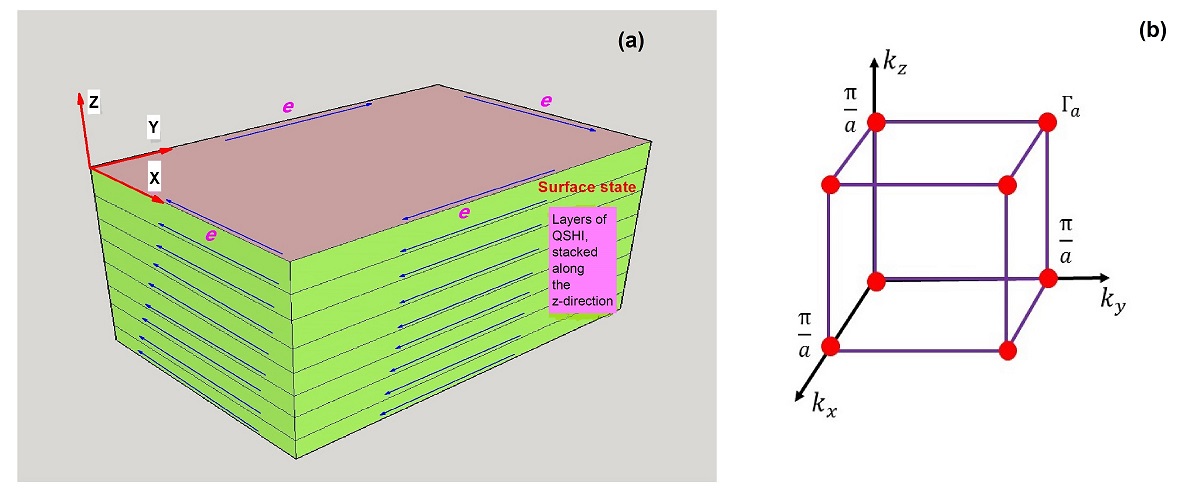}
\caption{(a) Weak three-dimensional topological insulators. (b) High-symmetry points in 3D bulk Brillouin zone.}\label{rh}
\end{figure}
\subsection{Strong topological insulators}\label{PO}
\begin{figure}[h]
\includegraphics[width=1.0\textwidth]{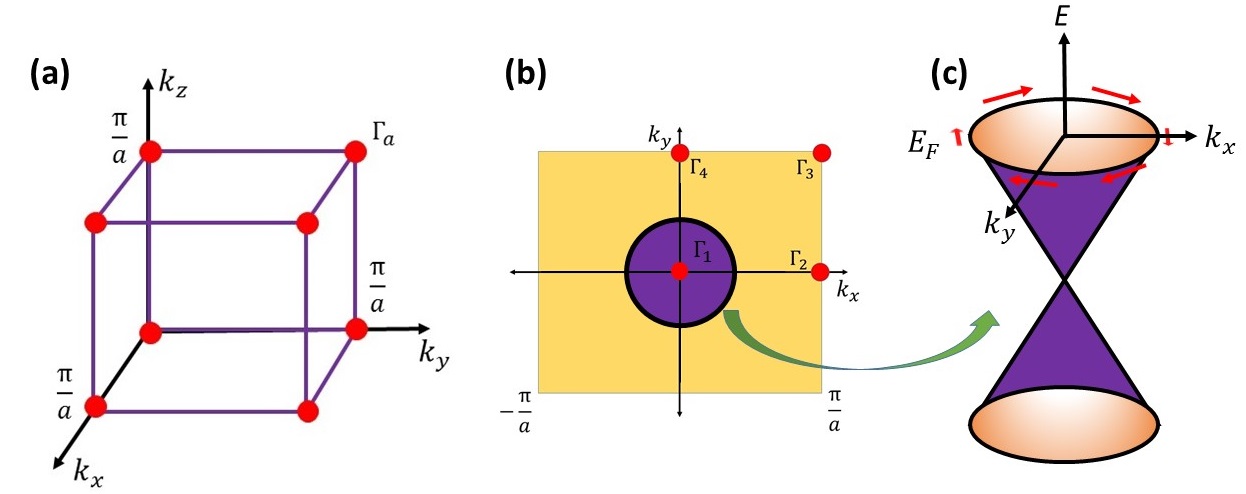}
\caption{(a) High-symmetry points in 3D bulk Brillouin zone. (b) Constant energy contour at the Fermi level. (c) 2D Dirac cone surface state and spin-momentum locking. }\label{rh}
\end{figure}
The strong topological invariant, $\nu_0$, for a three-dimensional bulk insulating state is determined by the expression,
\begin{equation}
( -1) ^{\nu_0} = \prod_{a = 1}^{8} \delta_a,
\end{equation}
where `a' is the index of time-reversal invariant points [Figure 10(a)] of bulk Brillouin zone. The materials, where the value of $\nu_0$ is found to be one, are known as strong topological insulators. As all the eight time-reversal invariant points are involved in determining the value of $\nu_0$, strong TI state cannot be interpreted as a descendant of the 2D-QSHI. The surface Brillouin zone, as shown in Figure 10(b), consists of four time-reversal invariant points, where the surface state must be Kramers degenerate. Away from these special points, the spin-orbit interaction lifts the degeneracy. As discussed in Section 1.3.2, for non-trivial surface state, the surface band structure must resemble the situation in Figure 7(b). By looking the constant energy contour of the Fermi level [Figure 10(b)], one can see that the surface Fermi circle encloses odd number of time-reversal invariant points for strong 3D TI. The novelty of this conducting surface state is the rich physics associated with the electronic band dispersion \cite{2,3}. It has been found that the dynamics of charge carriers on the surface of a 3D TI is governed by the Dirac-type effective Hamiltonian,
\begin{equation}
H_{sur}(k_{x},k_{y}) = -\hbar v_{F}\hat{z}\times\vec{\sigma}.\vec{k}.
\end{equation}
As a consequence, the energy and momentum of charge carriers follow gapless linear dispersion [Figure 10(c)], unlike conventional electronic system, where the dispersion relation is quadratic in nature. Another interesting characteristic of 3D TI surface state is that the spin of the charge carriers is always perpendicular to its momentum direction, known as spin-momentum locking. This makes the motion of charge carriers robust, against the non-magnetic impurity in a sample. This can be inferred from a simple logic. If there is any non-magnetic impurity in the system, to change the direction of motion of charge carrier, the impurity has to flip the direction of the spin. However, a scalar field (impurity potential) cannot affect a vector field (spin). So, there will be no backscattering of charge carriers.\\
\begin{figure}[h]
\includegraphics[width=1.0\textwidth]{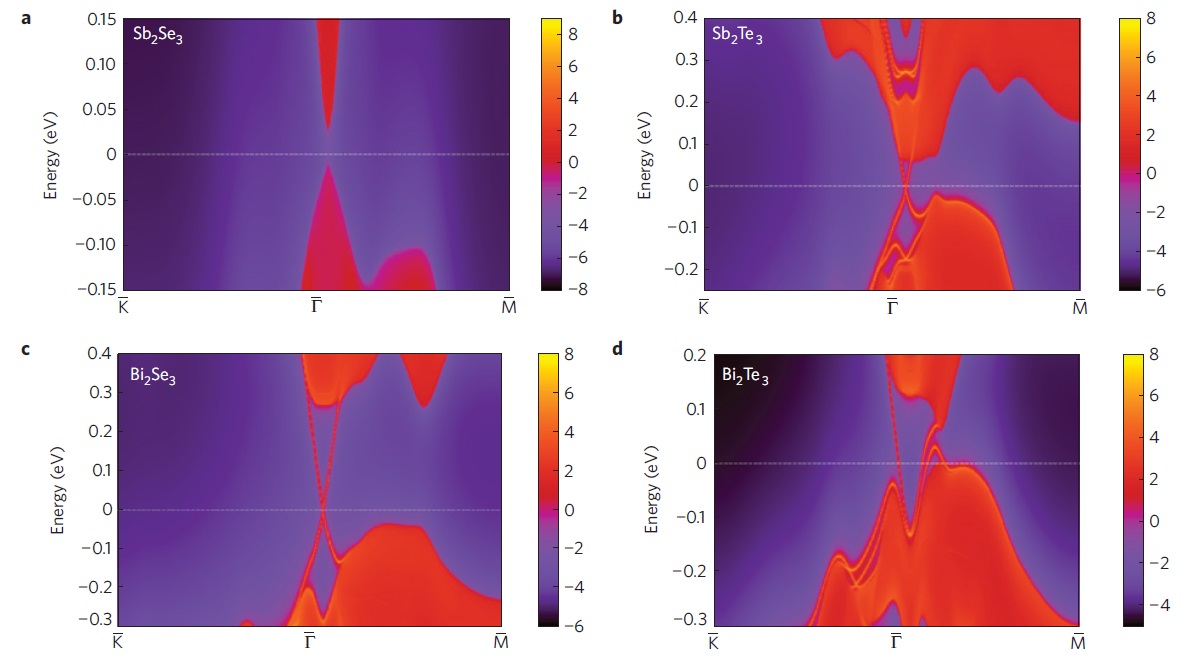}
\caption{(a)-(d) Energy and momentum dependence of the local density of states for the Bi$_{2}$Se$_{3}$ family of materials on the [111] surface. A warmer color represents a higher local density of states. Red regions indicate bulk energy bands and blue regions indicate a bulk energy gap. The surface states can be clearly seen around the $\Gamma$ point as red lines dispersing inside the bulk gap. Reproduced from Ref. \cite{18}.}\label{rh}
\end{figure}
Following the specific prediction of Fu and Kane \cite{16}, the 3D TI state has been first experimentally identified in Bi$_{0.9}$Sb$_{0.1}$ by Princeton University group led by Hasan, through the angle-resolved photoemission spectroscopy (ARPES) experiment \cite{17}. This material is an alloy of Bi and Sb, which possesses two essential features: (i) band inversion at odd number of time-reversal invariant momentum points in the bulk Brillouin zone, and (ii) opening of band gap at these points. This leads to non-trivial bulk $\mathbb{Z}_{2}$ topological invariant, which has been identified as (1; 1 1 1). The surface electronic band structure of this compound has been found to be complicated and the bulk band has a small insulating gap. As a consequence, at finite temperature, due to presence of thermally excited carriers, the quadratic bulk band has significant contribution in electronic transport. However, the overwhelming goal in the research on 3D TI is the realization of transport properties associated with the conducting surface state and utilization of this in next generation electronic device. To achieve this goal, it is necessary to find new materials, which have single spin-momentum locked Dirac cone surface state and large insulating gap in the bulk. Zhang \emph{et al.} came up with a concrete prediction that Bi$_{2}$Se$_{3}$, Bi$_{2}$Te$_{3}$, and Sb$_{2}$Te$_{3}$ are 3D TIs but Sb$_{2}$Se$_{3}$ is not \cite{18}. The electronic band structures of these materials, containing isolated surface and bulk states, are shown in Figure 11(a)-(d). Experimentally, the existence of a single Dirac-cone surface state was reported in 2009 for Bi$_{2}$Se$_{3}$ by Xia \emph{et al.} \cite{19}, for Bi$_{2}$Te$_{3}$ by Chen \emph{et al.}\cite{20} and also by Hsieh \emph{et al.} \cite{21}, and for Sb$_{2}$Te$_{3}$ by Jiang \emph{et al.} \cite{22}.\\
\section{Three-dimensional Dirac semimetal and its derivatives (e.g. Weyl semimetal)}\label{PO}
The research on TI and the experimental discovery of graphene band structure [Figure 12] have triggered a tremendous interest in condensed matter physics, over the past decade. The energy-momentum dispersion of charge carrier as well as the form of the underlying Hamiltonian for the surface state of 3D TI and in the bulk of graphene are reminiscent of those for massless fermions, usually studied in high-energy physics, with two relevant differences. First, the characteristic velocity that appears in condensed matter physics is roughly two orders of magnitude smaller than the speed of light. And second, both in graphene and 3D TIs, the electrons are constrained to move in two spatial dimensions, whereas the framework of relativistic quantum mechanics was established to describe fermions in three spatial dimensions. However, the constant efforts for the realization of relativistic particles in table top experiments result in new quantum phases of matter, which have linear dispersion along all the three momentum ($k_x$, $k_y$, $k_z$) directions in the shape of a cone. The materials, which host this type of electronic band structure are known as 3D Dirac semimetals.\\
\begin{figure}[h]
\includegraphics[width=1.0\textwidth]{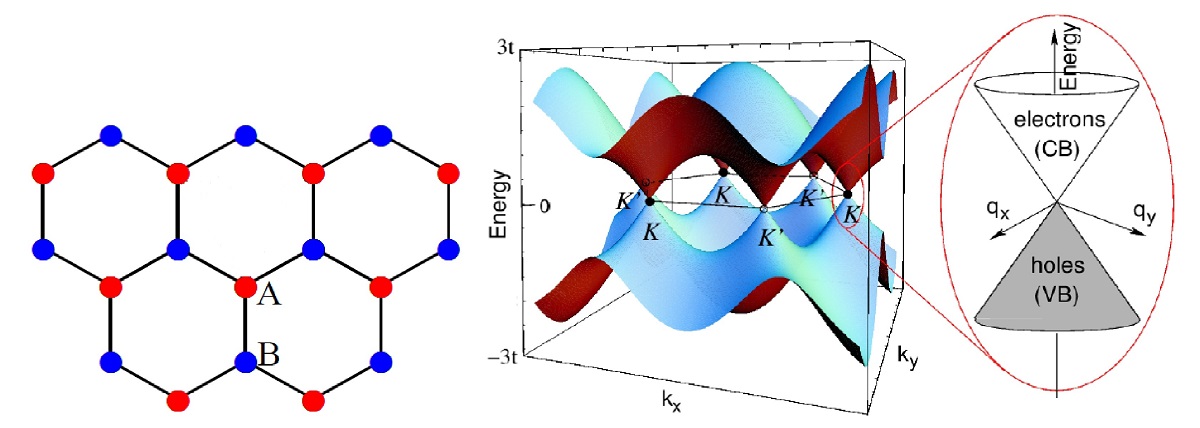}
\caption{Left: Honeycomb lattice of graphene. \textbf{A} and \textbf{B} are the two types of lattice, identified for band structure calculations, using tight-binding model. Right: Energy bands of graphene obtained from the tight-binding model and zoom around the Dirac point at K. Reproduced from Ref. \cite{23}.}\label{rh}
\end{figure}
\subsection{3D Dirac semimetal state at quantum critical point}\label{PO}
\begin{figure}[h]
\includegraphics[width=1.0\textwidth]{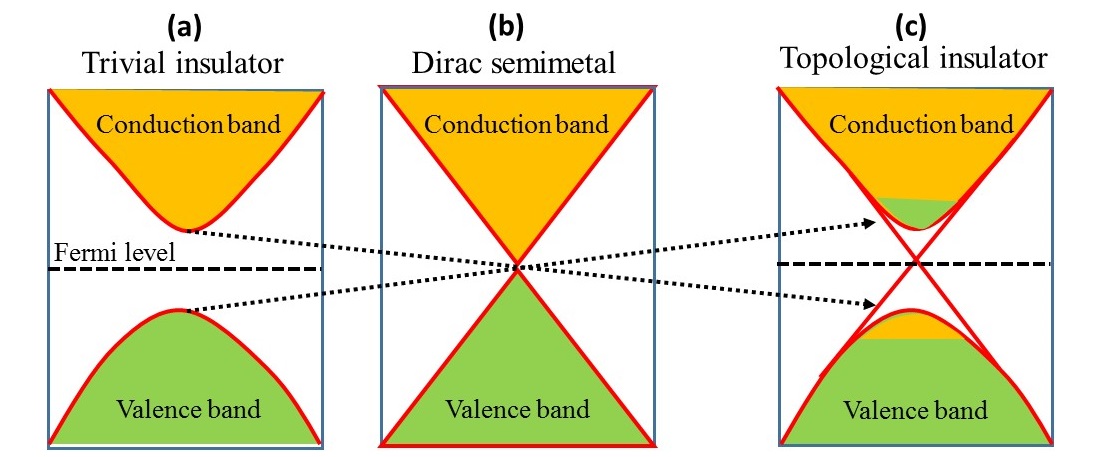}
\caption{Schematic band inversion between two bands: The trivial band gap in (a) closes at a critical point in (b), and reopens inverted in (c) with the two states swapping their orbital characters at the symmetry point.}\label{rh}
\end{figure}
It has been predicted that 3D Dirac semimetal state can be realized at a quantum critical point in the phase transition from a trivial insulator to a topological insulator \cite{8,16}. In an insulating material, the bulk band gap can be tuned by chemical doping or external pressure, which actually changes the lattice parameters and spin-orbit coupling in the system. This type of physical operation can even change the parity of an insulating gap from trivial to non-trivial, and vice versa. In the process of band evolution, the insulating gap for an inversion symmetric crystal has been found to be zero at some unique value of tuning parameter. At this critical value, the bulk conduction and valence bands touch at a special point in momentum space (which is known as `Dirac node'), and the dynamics of quasi-particles in the bulk electronic band of the material is governed by the Dirac-type equation for massless fermion in three dimensions,
\begin{equation}
i\frac{d\Psi}{dt} = H \Psi = \hbar V_F \left(
\begin{array}{cc}
 \overrightarrow{\sigma}.\overrightarrow{k} & 0 \\
 0 & -\overrightarrow{\sigma}.\overrightarrow{k} \\
 \end{array}
\right)\Psi.
\end{equation}
Here, $\overrightarrow{\sigma}$, $\overrightarrow{k}$ and $V_F$ are the Pauli spinor, crystal momentum, and Fermi velocity of charge carriers, respectively. In solid state crystallographic environment, speed of light ($c$) and linear momentum ($\overrightarrow{p}$) of original Dirac equation are replaced by $V_F$ and $\overrightarrow{k}$, respectively. As the Pauli matrices are two-dimensional, $H$ is a 4$\times$4 matrix, and the Equation 1.8 has four components. Following the theoretical prediction \cite{16,24}, the 3D Dirac semimetal state has been naively identified in Bi$_{1-x}$Sb$_{x}$ at a quantum critical point x = 0.04, through ARPES experiment \cite{17}. Later on, similar topological phase transition has been observed in BiTl(S$_{1 - \delta}$Se$_{\delta}$)$_{2}$, which is shown in Figure 14 \cite{25}. With increasing selenium concentration, the direct bulk band gap reduces from 0.15 eV at $\delta = 0.0$ to 0.05 eV at $\delta = 0.4$. At $\delta = 0.6$, the bulk conduction and valence bands touch each other, resulting a 3D Dirac dispersion. For compositions $\delta \geq 0.6$, the material becomes an inverted indirect band gap insulator with spin-polarized topological surface state. However, Bi$_{1-x}$Sb$_{x}$ and BiTl(S$_{1 - \delta}$Se$_{\delta}$)$_{2}$ fail to create significant excitement due to some limitations. As the Dirac cone state appears at a particular chemical composition in these compounds, it is not robust against uncontrolled doping during sample preparation. In addition, it has been found that the presence of conventional quadratic band masks the non-trivial band, which undergoes topological phase transition with chemical doping.\\
\begin{figure}[h]
\includegraphics[width=1.0\textwidth]{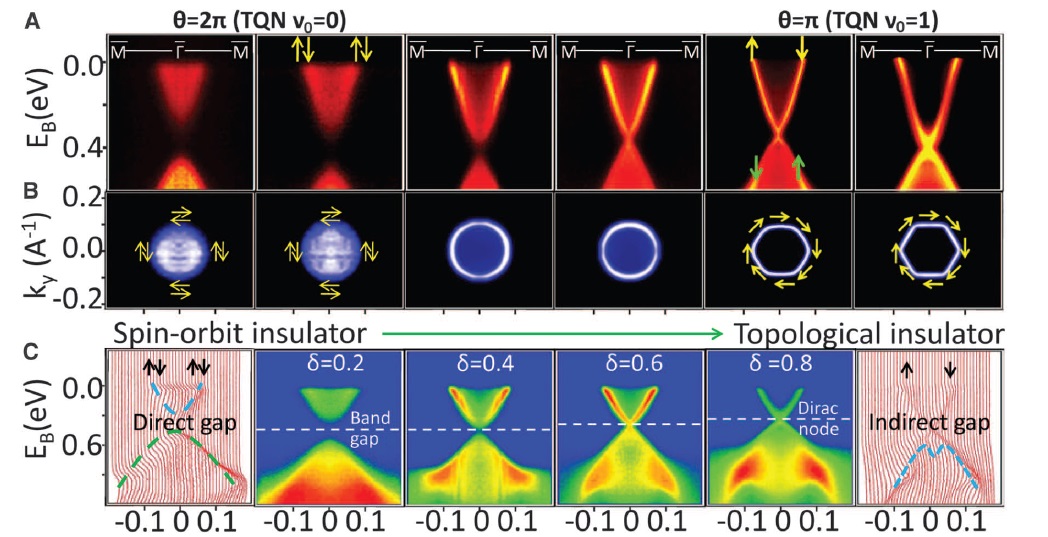}
\caption{Topological phase transition in BiTl(S$_{1 - \delta}$Se$_{\delta}$)$_{2}$. (A) High-resolution ARPES dispersion maps from a spin-orbit band insulator
(left panel) to a topological insulator (right panel). Topological quantum numbers (TQN) are denoted by topological invariant $\nu_0$. (B) ARPES-mapped native Fermi surfaces and their spin-texture for different chemical compositions (from left to right, $\delta = 0.0$ to $\delta = 1.0$). (C) Left and right panels: Energy distribution curves for stoichiometric compositions $\delta = 0.0$ and $\delta = 1.0$, respectively. Center panels: ARPES spectra indicating band gap and Dirac node for compositions $\delta = 0.2$ to $\delta = 0.8$. Figures are adopted from Ref. \cite{25}.}\label{rh}
\end{figure}
\subsection{Crystalline symmetry protected 3D Dirac semimetal}\label{PO}
It is important to note that the previously discussed scenario of topological phase transition and the emergence of Dirac cone state at the quantum critical point do not take into account any additional space group symmetries, which, if present may alter the conclusion \cite{8}. Several theoretical studies have predicted the existence of second generation 3D Dirac semimetals, where the Dirac cone band appears from the protection of certain space group crystalline symmetries, and are, therefore, proposed to be more robust to disorders or chemical alloying \cite{26,27,28}. For example, theoretical studies have identified Na$_{3}$Bi and Cd$_{3}$As$_{2}$ as 3D Dirac semimetals, which are protected by the C$_{3}$ and C$_{4}$ crystalline rotational symmetry, respectively \cite{27,28}. This type of Dirac semimetal, which is also known as 3D topological Dirac semimetal (TDSM), differs from the earlier-mentioned type because it possesses strong spin-orbit coupling driven inverted bulk band structure. At the special momentum points along the  symmetry axis, the band crossings are protected by the space group symmetry. Since both time-reversal and inversion symmetries are present, there is a fourfold degeneracy at these points, around which the band dispersions can be linearized, resulting in a 3D massless Dirac semimetal. The C$_{4}$ rotational symmetry protected unavoidable band crossing in Cd$_{3}$As$_{2}$ is shown in Figure 15(a), as a representative. The surface state of TDSM is also distinct from the closed constant energy contour in 3D TIs [Figure 15(b)], identified as Fermi arc. As shown in Figure 15(c), spin-momentum locked arc-like contour at the Fermi level connects two discrete points in surface Brillouin zone, which are the projection of bulk Dirac nodes on the surface. There is another important difference between the surface state of 3D TI and TDSM in the spin texture on the constant energy contour. The magnitude of spin projection perpendicular to momentum directions is constant throughout the closed loop in case of 3D TI. Whereas, in TDSM, it gradually tends to zero as the Fermi-arc approaches towards the points of discontinuity.\\
\begin{figure}[h]
\includegraphics[width=1.0\textwidth]{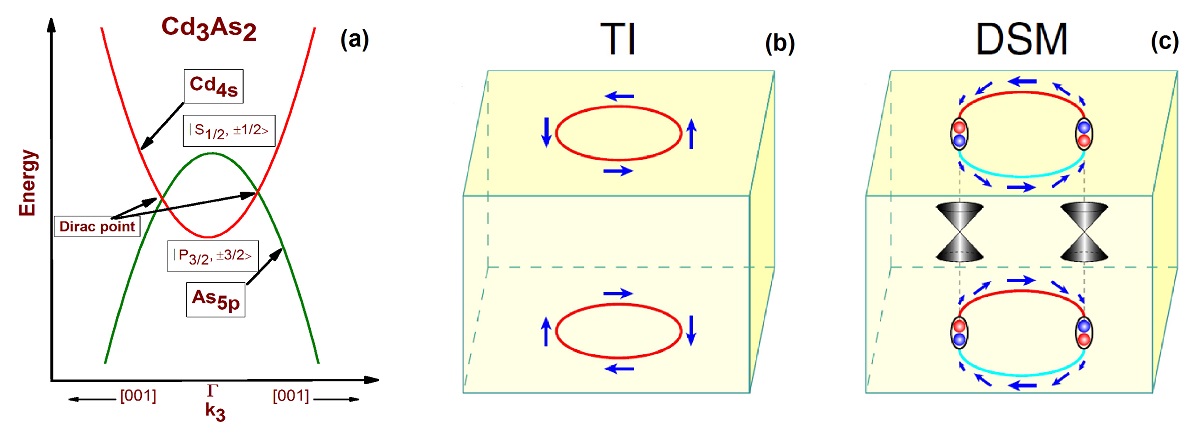}
\caption{(a) C$_{4}$ rotational symmetry protected unavoidable band crossing in Cd$_{3}$As$_{2}$, as a representative. Here, k$_{3}$ is the third momentum direction, i.e., k$_{3}$ direction. (b) Schematic of the spin-polarized surface states in a 3D TI. (c) Schematic of a TDSM with spin-polarized Fermi arcs on its surface connecting projections of two bulk Dirac nodes. The red and blue balls, surrounded by a black boundary indicates that one Dirac node is the degeneracy of two Weyl nodes, which will be discussed in the following section. Figure (b) and (c) are reproduced from Ref. \cite{29}.}\label{rh}
\end{figure}
Understanding the dynamics of relativistic Dirac fermions in table-top experiments is not the only fundamental importance of TDSM phase in solid state electronic systems. It has been predicted that by breaking different symmetries of a crystal, having this novel electronic phase, different new quantum phases of matter can be observed \cite{27,28}. It has been theoretically understood that under broken time-reversal symmetry scenario, in external magnetic field or upon magnetic impurity doping, TDSM acts like a 3D Topological Weyl semimetal (TWSM). Breaking the inversion symmetry of a TDSM, 3D TI state or TWSM state can be induced, depending on the crystalline space group symmetry of the material. By tuning the space group symmetry of a TDSM, axion insulator state can be induced. Chemical doping can also lead to new exotic phases such as, topological superconductivity.  We will provide a brief overview on TWSM state of matter, before going to discuss the experimental discovery of space group symmetry protected Dirac semimetals and the recent advancement in experiment.\\
\subsection{Understanding Topological Weyl semimetal as a transmuted state of Topological Dirac semimetal}\label{PO}
\begin{figure}[h]
\includegraphics[width=1.0\textwidth]{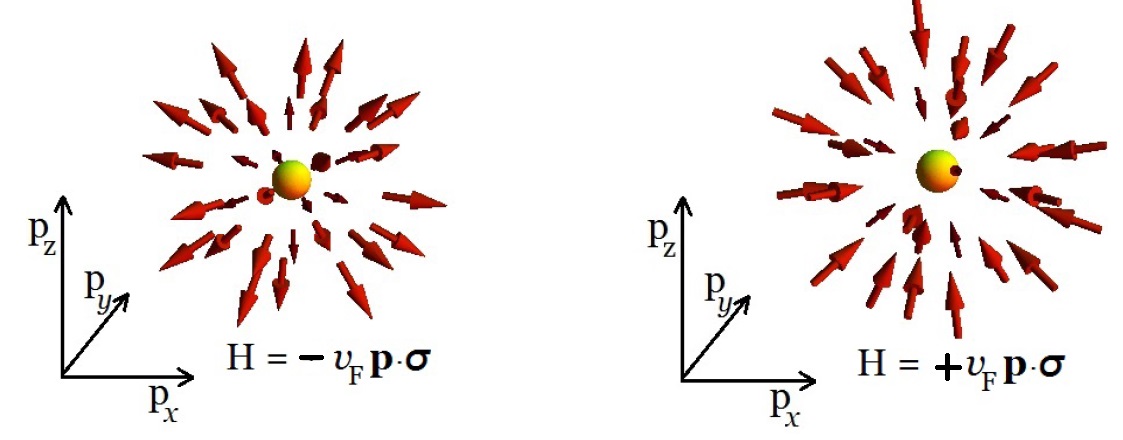}
\caption{Weyl nodes of opposite chirality. The arrows indicate the direction of the spin vector, which can be parallel or antiparallel to the momentum vector.}\label{rh}
\end{figure}

\begin{figure}[h]
\includegraphics[width=1.0\textwidth]{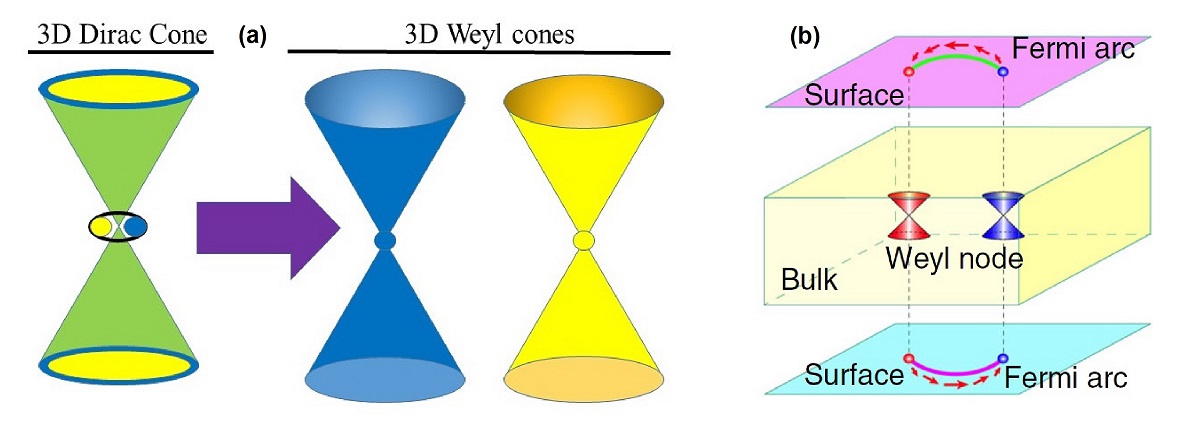}
\caption{(a) A four-component 3D Dirac node in a TDSM as a superposition of two two-component Weyl nodes, and the splitting of Dirac cone into two Weyl cones of opposite chirality under broken time-reversal symmetry.  (b) Schematic of a WSM with spin-polarized Fermi arcs on its surfaces connecting the projections of two Weyl nodes with opposite chirality. The red and blue colors of the bulk Weyl cones and the corresponding projection points on the surfaces represent opposite chirality of the Weyl nodes. The red arrows on the surfaces indicate the spin texture of the Fermi arcs. Figure (b) is reproduced from Ref. \cite{29}.}\label{rh}
\end{figure}
In the year 1929, Hermann Weyl proposed that a four-component massless Dirac equation [Equation 1.8] in three dimensions can be separated into two two-component equations \cite{30},
\begin{equation}
i\frac{d\Psi}{dt} = H \Psi = \pm c \overrightarrow{\sigma}.\overrightarrow{p}\Psi.
\end{equation}
The above equation describes particles with a definite projection of spin to its momentum, known as Weyl fermions. When the sign on the right hand side of the equation is positive, $\overrightarrow{\sigma}$ has to be antiparallel to $\overrightarrow{p}$, to minimize the energy. Massless fermions, obeying this specific spin orientation, are the positive chirality Weyl fermions. Again for the particles, which obey the above-mentioned equation with the negative sign, the spin has to be parallel to the momentum direction. This type of particles are called negative chirality Weyl fermions. If we look at the momentum space [Figure 16], it will be found that the expectation value of $\overrightarrow{\sigma}$ in an eigenstate of a given chirality forms a vector field, like a hedgehog. In condensed-matter physics, specifically in solid-state band structures, Weyl fermions appear when two electronic bands cross and low energy effective Hamiltonian around the band crossing point mimics the expression, $H = \pm c \overrightarrow{\sigma}.\overrightarrow{p}$. The crossing point is called a Weyl node, away from which the bands disperse linearly in the lattice momentum, giving rise to 3D Weyl semimetal state. As illustrated in Figure 17(a), the TWSM state can also be generated by breaking the time-reversal symmetry or inversion symmetry of a 3D Dirac semimetal, where a single four-component Dirac cone splits into two two-component Weyl cones. Theory also predicts that the materials, which possess Weyl fermions in the bulk electronic state, would exhibit a new kind of surface state: an open Fermi arc that would connect two Weyl nodes and then continue on the opposite surface of the material [Figure 17(b)] \cite{31}. In the year 2015, two groups simultaneously predicted the existence of Weyl-type electronic excitations in TaAs, TaP, NbAs, and NbP \cite{32,33}. Subsequent after the theoretical predictions, the first experimental discovery of Weyl semimetal state in TaAs family of materials has been done by Xu \emph{et al.} \cite{34} and Lv \emph{et al.} \cite{35}.

\subsection{Experimental discovery of Topological Dirac Semimetal}\label{PO}
\begin{figure}[h]
\includegraphics[width=1.0\textwidth]{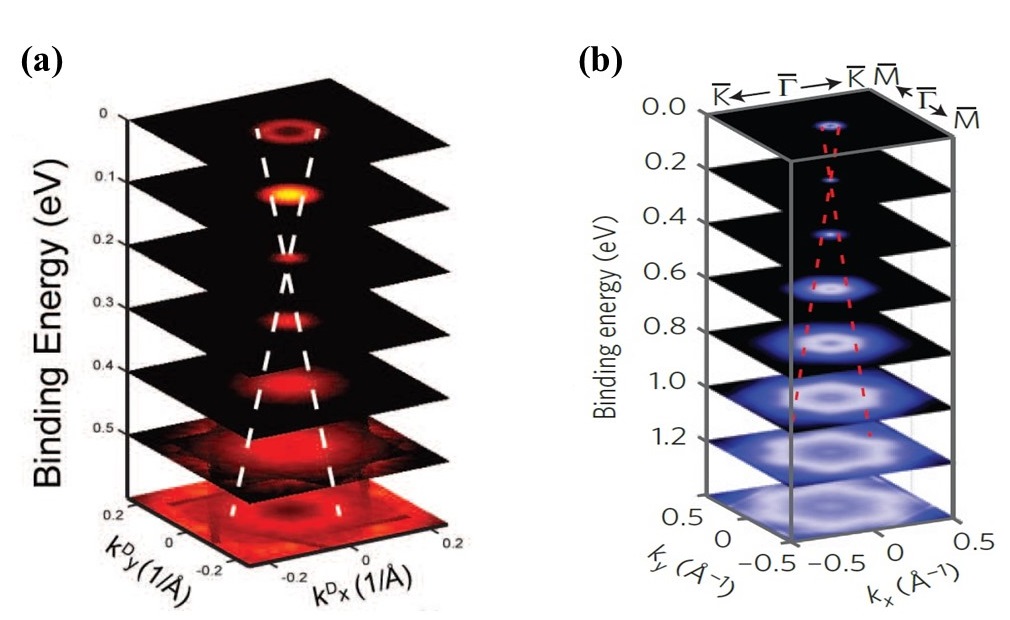}
\caption{Stacking plot of constant-energy contours at different binding energies shows Dirac cone band structure. (a) White dashed lines are the guide to the eye that trace the linear dispersions in Na$_{3}$Bi. Figure reproduced from Ref. \cite{36}. (b) Red dotted lines are guide to the eye that indicate the linear dispersions and intersect at the Dirac point in Cd$_{3}$As$_{2}$. Figure reproduced from Ref. \cite{37}}\label{rh}
\end{figure}
Following the theoretical prediction \cite{27,28}, investigation on electronic band structure of Na$_{3}$Bi and Cd$_{3}$As$_{2}$ through ARPES experiment have established the Dirac cone band dispersion in these compounds \cite{36,37}. Stacking plots of constant-energy contours at different binding energies for both the compounds are shown in Figure 18, where the gradually increasing radius of the circular contours lies on a straight line passing through the Dirac nodes. Immediate after the observation of bulk Dirac cone band, Yi \emph{et al.} and Xu \emph{et al.} have revealed the existence of Fermi-arc surface state in Cd$_{3}$As$_{2}$ and Na$_{3}$Bi \cite{38,39}. Although both the materials are equally compelling, experimental research on Na$_{3}$Bi has been found to be little challenging compared to Cd$_{3}$As$_{2}$ due to its extreme sensitivity to air. Later on, 3D topological Dirac semimetal phase was theoretically predicted and experimentally proposed in plenty of compounds. However, the existence of this novel electronic phase has been unambiguously established in a very few materials. ZrTe$_{5}$ and ZrSiS are the examples of materials, which have emerged as suitable candidates for extensive experimental research \cite{40,41,42}.\\

\subsection{Further classification of Topological Dirac and Weyl Semimetal}\label{PO}
\begin{figure}[h]
\includegraphics[width=1.0\textwidth]{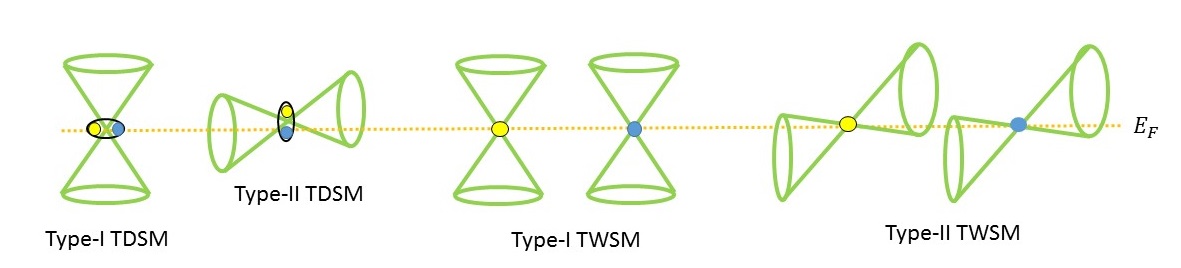}
\caption{Types of TDSM and TWSM, classified based on the band crossing. (a) Type-I Dirac Semimetal, (b) Type-II Dirac Semimetal, (c) Type-I Weyl semimetal, (d) Type-II weyl semimetal. }\label{rh}
\end{figure}
Topological Dirac and Weyl semimetal further classified into two categories. Depending on the orientation of band crossing in momentum space, these semimetals can be identified as type-I or type-II. For type-I topological semimetals, as shown in Figure 19(a) and (c), by tuning the chemical potential it is possible to avail the Dirac/Weyl node without introducing any finite density of states in the system. Whereas for type-II cases, as shown in Figure 19(b) and (d), tilted orientation of Dirac/Weyl cones force to introduce finite density of states at the Fermi level at any value of chemical potential Type-II Dirac/Weyl points always appear at the contact of electron and hole pockets. The fundamental principle of nature ensure that the laws of physics or any experimental results should be independent of the orientation or the boost velocity of the reference frame, known as Lorentz invariance. Form the earlier discussion, it is evident that type-II topological semimetals violetes Lorentz symmetry. All the materials mentioned in section C and D host type-I Dirac/Weyl band crossing in electronic structure. Later on, WTe$_{2}$, MoTe$_{2}$, etc. have been found to host type-II Weyl semimetal state \cite{43,44}. Type-II Dirac band crossing has been either proposed or identified in VAl$_{3}$ and PtTe$_{2}$ \cite{45,46}.

\section{Topological semimetal beyond Dirac and Weyl}
In high-energy physics, the relativistic fermions are protected by Poincare symmetry (i.e., translation + Lorentz symmetry), while in condensed matter, they respect one of the 230 space group symmetries. The variation of crystal symmetry from one material to another escalates the potential to explore free fermionic excitations such as Dirac, Weyl, Majorana and beyond. Subsequent after the discovery of Dirac nd Weyl semimetal, Bradlyn et al. have predicted the existence of exotic fermions near the Fermi level in several materials, governed by their respective space group symmetry \cite{47}. Unlike two- and four-fold degeneracy in Weyl and Dirac semimetals, these systems exhibit topology protected three-, six- , and eight-fold degenerate band crossing at high symmetry points in the Brillouin zone. For example, three-component fermion has recently been observed in molybdenum phosphide \cite{48} and tungsten carbide \cite{49}. Bradlyn et al. have suggested many materials, and many of them have reported to exist in single crystal form such as,  Ag$_{3}$Se$_{2}$Au, Pd$_{3}$Bi$_{2}$S$_{2}$, LaPd$_{3}$S$_{4}$, Li$_{2}$Pd$_{3}$B, and Ta$_{3}$Sb. A schematic of three-component band crossing and its difference from Dirac and Weyl cone have been shown in Figure 20(a). Crossing between two doubly-degenerate bands, two non-degenerate bands, and one doubly degenerate and one non-degenerate bands, leads to four-fold degenerate Dirac fermion, two-fold degenerate Weyl fermion and three-component fermion, respectively. It is to be noted that not only one doubly spin degenerate band and one spin non-degenerate band produce three-component fermion but also the band crossing between three spin non-degenerate bands leads to three-component fermion. An example has been shown in Figure 20(b) for Pd$_{3}$Bi$_{2}$S$_{2}$. Similar to Dirac and Weyl semimetals, these three-fold or higher-fold degenerate topological semimetals also have Fermi arc surface state \cite{49}.
\begin{figure}[h]
\includegraphics[width=1.0\textwidth]{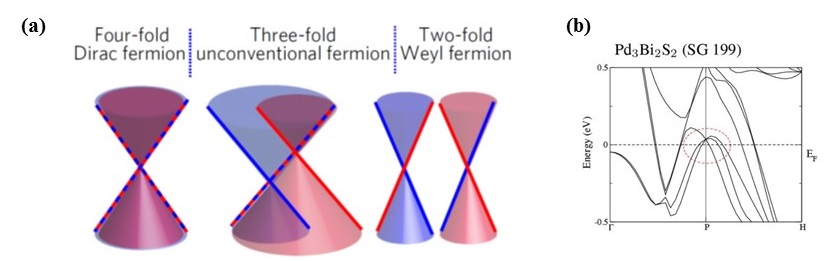}
\caption{Schematic and band structure of materials beyond Dirac and Weyl semimetals. (a) Schematic of three-component/three-fold Fermion. Each colour represents band of a particular spin type. Figure reproduced from Ref. \cite{49}. (b) Three-fold band crossing in Pd$_{3}$Bi$_{2}$S$_{2}$, as predicted in Ref. \cite{47}.}\label{rh}
\end{figure}

\section{Topological nodal line semimetal}\label{PO}
In three-dimension there is another class of topological semimetal, where the conduction and valence bands cross each other along a line, unlike at a discrete point in TDSM or TWSM. Materials which host this type of band crossing are known as topological nodal line semimetal (TNLSM). The nodal line can be either closed loop or a discrete line. The schematic of nodal line band crossings and its contrast with Dirac/Weyl node, have been shown in Figure 21. Electronically, TNLSM can be considered as an intermediate state of topological point node semimetals (i.e., Dirac/Weyl semimetals) and normal metals. The reasons behind that are the following: (i) exactly at the half filling, the Fermi surface of TDSM/TWSM, TNLSM, and normal metal are zero-, one-, and two-dimensional, respectively, and (ii) the density of states scale as $\rho_{0} \propto (E-E_{f})^{2}$, $\rho_{0} \propto |(E-E_{f})|$, and $\rho_{0} \propto \sqrt{E-E_{f}}$ in TDSM, TNLSM, and normal metal, respectively \cite{50}.\\

\begin{figure}[h]
\includegraphics[width=0.7\textwidth]{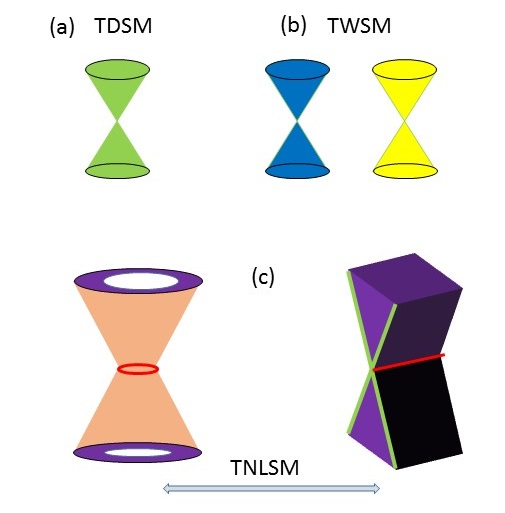}
\caption{Distinct features of topological nodal line. (a) Dirac node, (b) Weyl node, (c) Nodal line either in the form of closed loop or in the form of discrete line.}\label{rh}
\end{figure}
To protect nontrivial band crossing over a line, nodal line semimetal needs more no of symmetries and conditions compared to Dirac and Weyl semimetals. A combination of inversion plus time-reversal symmetry, mirror reflection symmetry,and non-symmorphic symmetry, and usually the absence of spin-orbit coupling (SOC) (there are few examples of TNLSM state in presence of SOC) require to protect the line node. If there is a nodal line state, which host Dirac-type spinless fermions as quasi-particle excitation, it can be tuned into different other topological states by breaking or imposing different symmetries and in presence of spin-orbit coupling \cite{51}. Starting from nodal line semimetal with spinless Dirac-type quasi-particle excitation, sufficient SOC could lift the degeneracy along the band crossing line and destroy the line node. However if inversion, time-reversal, and non-symmorphic symmetry are preserved in the system, it protect the line of degeneracy inspite of the fact that the spin degeneracy between the crossing bands is lifted. Similarly, different combinations of time-reversal, inversion, mirror, and rotation symmetry lead to nodal line semimetal with Weyl-type quasi-particle excitations, Dirac semimetal, Weyl semimetal, and topological insulator.  Schematic of different topological states and their relationship with Dirac-type spinless nodal line semimetal have been illustrated in Figure 22(a). Review article by S.-Y. Yanga et al. \cite{51}, discussed the above-mentioned topological electronic transformation in details.\\
\begin{figure}[h]
\includegraphics[width=1.0\textwidth]{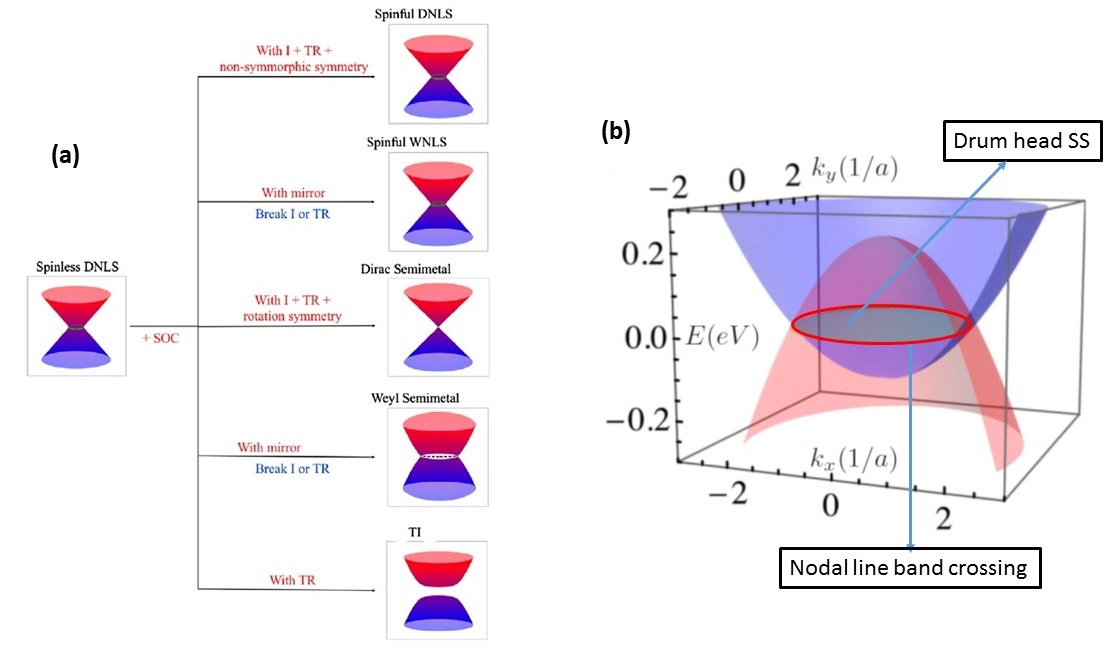}
\caption{Daughter states from spinless TNLSM and its surface state. (a) Different topological electronic states which can be realized from  TNLSM state upon breaking or imposing different symmetries and in presence of SOC. Red and blue indicate the symmetries which need to be preserved and broken, respectively. Adopted from Ref. \cite{51}. (b) Nodal line band crossing and drum head surface state. Adopted from Ref. \cite{52}}\label{rh}
\end{figure}

Unlike 1D Fermi arc surface state in Dirac/Weyl semimetals, nodal line semimetals host drumhead 2D surface state, as shown in Figure 22(b). In the three-dimensional Brillouin zone the 'drumhead’ surface state is embedded inside the ‘direct gap’ between conduction and valence band in the 2D projection of the nodal ring. This unique surface state is nearly dispersionless, analogous to the acoustic vibration on the surface of a drum, and gives rise to large density of states over the region \cite{51}. ZrSiS and PbTaSe$_{2}$ are among the first batch of proposed candidate materials which have been experimentally realized as Dirac nodal line semimetals and Weyl nodal line semimetal, respectively \cite{53,54}.\\

\section{Topological Crystalline Insulators (TCI)}\label{PO}
\subsection{A sense of new topological invariant from geometry}\label{PO}
Are Z (Integer quantum Hall) and Z$_{2}$ only topological classification of insulators in materials depending on the presence and absence of time reversal symmetry and dimensionality? The simple answer is, No. It is possible to found different topological classes or define different topological invariants on different characteristics of a system. In the following section, we have given an example from Geometry and then, we introduced topological insulators having topological invariant other than Z and Z$_{2}$. So far, we discussed 'Genus' in section II.B as a topological invariant in Geometry. However, `Euler characteristic' is another useful topological invariant in Geometry \cite{5}, condensed matter physicist know about. Before introducing the definition of 'Euler characteristic', we suggest the readers to imagine a polyhedron (K) in 3D space which is a geometrical object surrounded by faces. The boundary of two faces is an edge and two edges meet at a vertex. Two geometrical objects are considered to be equivalent (more correctly 'homeomorphic' in mathematical sense) if their `Euler characteristic', defined as $\chi$ = (number of vertices in K) - (number of edges in K) + (number of faces in K). So `Euler characteristic' is a number which can be used to classify polyhedrons in geometry. A numbers of geometrical object with their $\chi$ value have been shown in Figure 23.\\
\begin{figure}[h]
\includegraphics[width=1.0\textwidth]{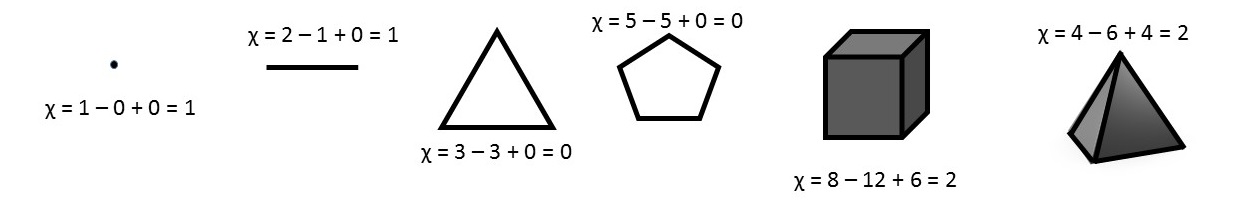}
\caption{Classification of Polyhedrons according to 'Euler characteristic', $\chi$.}\label{rh}
\end{figure}
\subsection{Crystalline topological invariant protected insulating state}\label{PO}
\begin{figure}[h]
\includegraphics[width=0.8\textwidth]{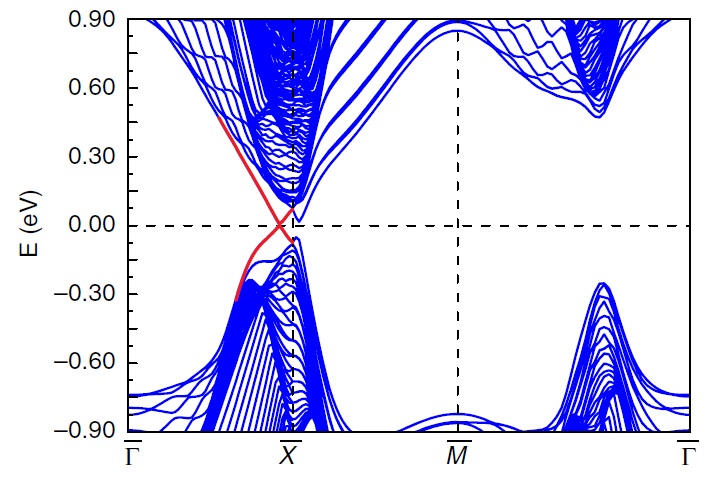}
\caption{A schematic of bulk band and surface state (red line) on (001) face of SnTe. Adopted from Ref. \cite{56}.}\label{rh}
\end{figure}
Similar to geometry, considering different point group symmetries of crystal (mirror, rotation, inversion, etc.), it is possible to introduce new topological invariant to characterize insulating bulk bands in materials \cite{55}. Metallic surface states which appear on high symmetry crystal surfaces, are similar to that observed in Z$_{2}$ TI (Figure 24). The gapless boundary states can only be gapped out and the bulk ground state can only be adiabatically connected to an atomic limit by breaking only the crystal symmetry, not the time-reversal symmetry as in case of Z$_{2}$ TI. The materials which are characterized in this way are known as 'Topological Crystalline Insulators (TCIs)'. For an example, in TCIs with mirror symmetry the topology is specified by the mirror Chern number, $n_{M}$. In this context, it will be worthy to mention that time-reversal symmetry can also be preserved in TCIs and it only add some fine structure to the gapless surface state. The first burst of excitation in this specific topic of research has been created by theoretical prediction of mirror symmetry protected TCI state in SnTe \cite{56}. From the band structure calculations, Hsieh et al. showed that SnTe has band inversions at four time-reversal invariant momentums (at four equivalent points in 3D Brillouin zone), which gives rise to a nontrivial mirror Chern number $n_{M}$ = , while its Z$_{2}$ invariant is trivial, (0;000). Subsequent after the theoretical prediction, gapless surface states inside bulk band gap and its spin-polarized Dirac cone structure have been observed through ARPES in SnTe and Pb doped SnTe \cite{57,58,59}. It is interesting to know that crystalline topological Chern number can be found in Z$_{2}$ TIs. For example, well-known Z$_{2}$ topological insulator Bi$_{2}$Se$_{3}$ has mirror Chern number, $n_{M}$ = -1. The sign of $n_{M}$ depends on the direction of spin-texture on constant energy contour of Surface Dirac cone band \cite{60}. Various values of mirror Chern number in different systems and its dependence on spin-texture orientation have been illustrated in Figure 25.

\begin{figure}[h]
\includegraphics[width=0.8\textwidth]{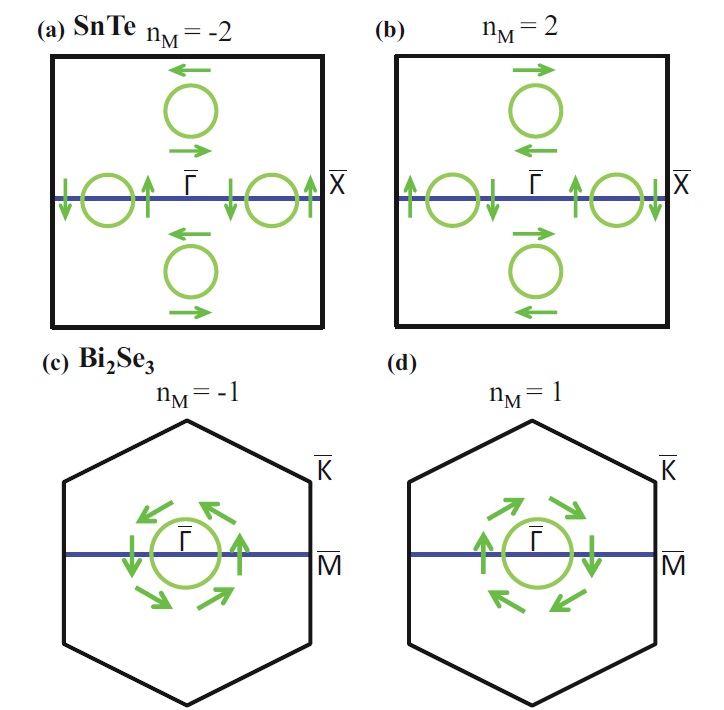}
\caption{Schematic diagrams for spin textures of the lower Dirac cones in distinct topological phases associated with various values of mirror Chern number. (a) $n_{M}$ = -2, (b) $n_{M}$ = 2, (c) $n_{M}$ = -1, and (d) $n_{M}$ = 1. Blue lines indicate the axis of mirror symmetry. adopted from Ref. \cite{60}.}\label{rh}
\end{figure}

\subsection{Higher order topological insulators}\label{PO}
So far we observed topological insulators in d dimensions have gapless states in (d-1) dimensional boundaries. According to the nomenclature introduced by F. Schindler et al. \cite{61}, these belong to first-order topological insulators. However, there exists three-dimensional topological insulators that host no gapless surface states, but exhibit topologically protected gapless states at hinge i.e., at (d-2) dimensional boundaries. This class of materials has been regarded as higher-order topological insulators. Their topological character is protected by combinations of specific crystal symmetries and time-reversal symmetry such as, time-reversal and a four-fold rotation symmetry, time-reversal and mirror symmetry. Second order TI state has been first realized in SnTe under small distortion induced by ferroelectric displacement along (111) direction \cite{61}.
\begin{figure}[h]
\includegraphics[width=0.8\textwidth]{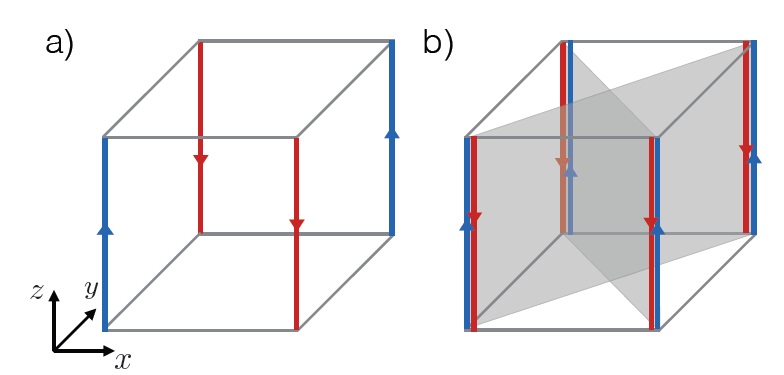}
\caption{Higher order topological insulators. (a) Time-reversal breaking model with $\hat{C}^{z}_{4}$-preserving bulk termination results in chiral hinge currents running along the corners. (b) Time-reversal invariant model with preserved mirror symmetries (Planes invariant under the mirror symmetries are highlighted in gray) results in anti-propagating Kramers pairs of hinge modes. Adopted from Ref. \cite{61}.}\label{rh}
\end{figure}

\section{Magnetic topological materials}\label{PO}
\begin{figure}[h]
\includegraphics[width=1.0\textwidth]{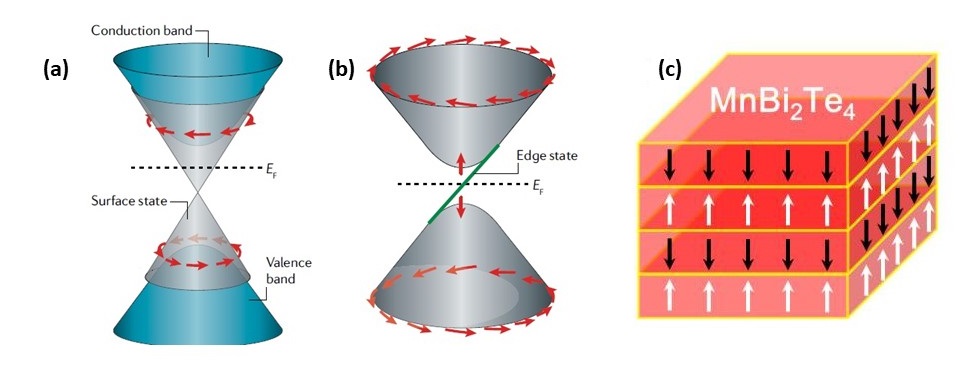}
\caption{Electronic structure of magnetic elements doped topological insulators and intrinsic AFM ordering in topological insulator MnBi$_{2}$Te$_{4}$. (a) The massless Dirac-like dispersion of the surface state with
spin-momentum locking in a topological insulator. The surface state band connects the bulk valence and the bulk conduction bands.(b) The gapped Dirac-like dispersion of the surface state in a magnetic topological insulator. The blue solid line inside the gap represents the band dispersion due to induced Quantum Anomalous Hall state at the edge. Figures (a) and (b) adopted from Ref. \cite{66}. (c) Interlayer magnetic ordering in vdW topological insulator, MnBi$_{2}$Te$_{4}$. Adopted from Ref. \cite{67}.}\label{rh}
\end{figure}
So far, presence of magnetism in a material did not add new topological class over the existing paradigm of topological materials. Magnetism actually plays the role of a new spice in the recipe of topological materials to make its electronic properties more exotic and exciting by breaking time-reversal symmetry or through the coupling with topological electronic bands. The magnetic-ion doping to 3D Z$_{2}$ TI has been started to see the effect of time-reversal symmetry breaking and as a result, the consequence of gap opening at the surface Dirac cone (Figure 27(a)-(b)). Examples are Mn or Fe or Cr doping in Bi$_{2}$Se$_{3}$, Bi$_{2}$Te$_{3}$, and Sb$_{2}$Te$_{3}$ \cite{62,63,64}. This was a topic of enormous research interest until recently, when topological insulating state with intrinsic anti-ferromagnetism (AFM) has been discovered in van der Waals (vdW) material, MnBi$_{2}$Te$_{4}$ \cite{65}. Considering interlayer AFM ordering in MnBi$_{2}$Te$_{4}$ (Figure 27(c)), it has been found that the combination of time-reversal and primitive-lattice translation symmetries introduced Z$_{2}$ topological classification of AFM insulators and the value of topological invariant is 1 for this material. This year a review article by Y. Tokura et al. has been published in Nature Review, where readers can folloe a detail history of magnetic topological insulators \cite{66}.\\
\begin{figure}[h]
\includegraphics[width=1.0\textwidth]{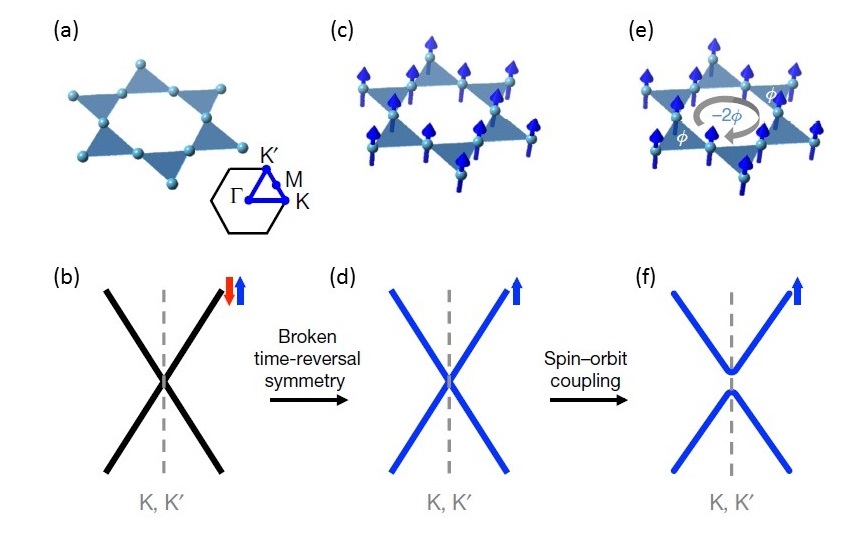}
\caption{The kagome structure and Fe$_{3}$Sn$_{2}$. (a) Structure of the kagome lattice, and (b) the associated Dirac point in the nearest-neighbour tight-binding model with the Brillouin zone shown in the inset. The
band is degenerate, as denoted with red and blue spins. (c) Ferromagnetic kagome lattice with broken time-reversal symmetry (moments in blue) and (d) the associated spin-polarized Dirac band with coupling between the magnetization and spin. (e) Spin-orbit-coupled ferromagnetic kagome lattice with Berry phase $\phi$ accrued via hopping and (f) the corresponding gapped Dirac spectrum.  Adopted from Ref. \cite{74}.}\label{rh}
\end{figure}
Similar to magnetic topological insulator, magnetism in topological semimetals is a subject of considerable research interest. As discussed, to be a Weyl semimetal either inversion or time-reversal or both have to be absent in a material. However, all the materials mentioned in section V.(c) belong to inversion symmetry breaking Weyl semimetal. The presence of magnetism is important to realize the spontaneous time-reversal symmetry breaking Weyl semimetal state. Although several ferromagnetic and anti-ferromagnetic materials have been theoretically proposed and experimentally established as time-reversal symmetry breaking Weyl semimetals, we have mentioned a few for examples such as, noncolinear antiferromagnet Mn$_{3}$Sn and Mn$_{3}$Ge \cite{68,69,70,71}, kagome-lattice ferromagnet Co$_{3}$Sn$_{2}$S$_{2}$ \cite{72}, and both ferromagnetic and antiferromagnetic ordered  BaMnSb$_{2}$ \cite{73}.
Not only in Weyl semimetal, presence of magnetism and its consequence on bulk Dirac cone band structure has also been understood concurrently. Ferromagnetic kagome metal Fe$_{3}$Sn$_{2}$ is the candidate material, where bulk Dirac band with mass gap $\sim$ 30 meV has been observed/realized through ARPES and scanning tunneling microscopy (STM) \cite{74,75}. Kagome lattice is a two-dimensional network of corner-sharing triangles, as shown in Figure 28(a). The simple tight-binding model in Kagome lattice structure predicts Dirac cone band in the bulk (Figure 28 (b)). However, the presence of ferromagnetic ordering (Figure 28 (c)) and as a consequence, the time-reversal symmetry breaking splits the spin-degenerate Dirac band (Figure 28 (d)). As illustrated in Figure 28 (e)-(f), further inclusion of spin-orbit coupling inject non-trivial Berry's curvature into the band structure and generate a mass gap in the Dirac cone. The Berry's phase accrued by the hopping of electrons is the source of intrinsic anomalous Hall effect in this material. In addition, magnetic frustration coming from corner sharing lattice and ferromagnetic order makes the material more exciting. A fascinating interplay between the massive Dirac band and magnetization directions, tuned by the external magnetic field, has been realized through STM and quantum oscillation study \cite{75,76}.\\

\textbf{Note:} \emph{In this review, we have not discussed topological superconductor and the Majorana physics due to its relatively vast and complex literature. We suggest the readers a review article by M. Sato et al. for reference \cite{77}. We might include this in future in revised version of this review.}

\end{document}